\documentclass[letter]{aa} 

\usepackage{graphicx}

\usepackage{txfonts}

\begin{document}

\title{Weighing Milky Way and Andromeda in an expanding $\Lambda$CDM Universe}
\subtitle{Decreasing the Local Group mass}

\author{David Benisty \inst{1,2}   }

\institute{Frankfurt Institute for Advanced Studies (FIAS), Ruth-Moufang-Strasse~1, 60438 Frankfurt am Main, Germany 
\and Helsinki Institute of Physics (HIP), P.O. Box 64, FI-00014 University of Helsinki, Finland  \\    \email{benidav@post.bgu.ac.il}}

   \date{}

 
\abstract
{The dynamics of the Local Group (LG), especially the contribution of the Milky Way (MW) and Andromeda (M31) galaxies, is sensitive to the presence of dark energy. This work analyzes the evolution of the LG by considering it as a two-body problem in a homogeneous and isotropic expanding spacetime in a full $\Lambda$ cold dark matter $(\Lambda$CDM) background. Using the timing argument (TA), which links LG dynamics to LG mass, we find that the complete $\Lambda$CDM background predicts a $\sim 10 \%$ lower mass for the LG; while $\Lambda$ alone predicts a $\sim 10 \%$ higher mass. The TA mass is modified by (i) simulations and (ii) the effect of the Large Magellanic Cloud (LMC) to alleviate the poorly constrained internal mass distributions of M31 and the MW, their time evolution, and the unknown distribution of dark matter between them. First, using IllustrisTNG simulations, we accounted for the effects of two extended halos and their environment (rather than point particles) and predicted their mass $\left(3.89 \pm 0.62\right) \cdot 10^{12} M_{\odot}$. Second, the LMC effectively changes the separation and velocities of M31 toward the MW and reduces the predicted mass to $\left(2.33 \pm 0.72\right) \cdot 10^{12} M_\odot$. Despite the uncertainties around dark matter between these galaxies, the overall estimated mass is compatible with the mere sum of the MW and M31 masses. The total mass of the TA is compatible with other estimates, such as the Hubble flow and the Virial Theorem with other dwarf galaxies. The combined result shows, for the first time, that a lower mass estimate can be obtained from the TA, with a consistent embedding and other systematic effects, and without an additional dark matter halo around the galaxies. }

\maketitle
%

\section{Introduction}
Dark energy is an unknown type of energy that influences the universe on its largest scales~\citep{Peebles:2002gy}. Beyond its impact on large scales, dark energy exerts a significant influence on the Local Universe~\citep{Chernin:2003qd,Chernin:2006dy,Peirani:2008qs,Chernin:2015nga,Karachentsev:2008st}. The cosmological constant ($\Lambda$) alters the predicted mass of the LG of galaxies, assuming that it consists mainly of MW and M31, which form a two-body system~\citep{chernin:2009a,Partridge:2013dsa}. An elegant approach to inferring the total mass of the LG is the timing argument (TA), which connects the observed kinematics of the LG to the LG mass~\citep{bib:Kahn,bib:Lynden,1989MNRAS.240..195R,Sawala:2023sec}. This argument assumes that the present system is on its first encounter in a two-body orbit that has been continuously expanding from the Big Bang until now. Initially, only assuming a radial velocity, the TA has since been extended to include eccentricity \citep{Li:2007eg} and the recoil velocity of the MW with respect to the LMC \citep{Chamberlain:2022fqr,Benisty:2022ive}. However, the estimated mass of the TA is upwardly biased  (about $[4-6]\cdot 10^{12}\,M_{\odot}$), compared to other estimates (see Section 2 and Fig.~(1) in~\cite{Sawala:2022ayk}, where an overview of different estimates is given).

A non-trivial effect of the LG is the repulsion effect of $\Lambda$. The TA mass is predicted to be about 10\% higher in the presence of $\Lambda$~\citep{Partridge:2013dsa} due to a constant repulsion force that needs to be compensated for by a higher mass and a higher Newtonian attraction force. This Letter demonstrates that including the complete history of $\Lambda$ cold dark matter ($\Lambda$ CDM) predicts a 10\% lower mass than the TA mass since, in earlier times, the embedding endowed it with a force of attraction. With simulation calibration and the effect of LMC, we show, for the first time, that it is possible to achieve a total mass of the LG $\left(2.33 \pm 0.72\right) \cdot 10^{12} M_{\odot}$, which is lower then the previous TA mass estimates.

The structure of the letter is as follows: Section~\ref{sec:emb} details the different embedding spacetime. Section~\ref{sec:effects} presents a discussion of the different effects included in estimating the mass (i) embedding spacetime, (ii) simulation calibration, and (iii) inclusion of the LMC. Section~(\ref{sec:lgmass}) gives the different mass estimates, along with the various related effects and a comparison of the masses with other masses in the literature.

\section{Embedding spacetime}
\label{sec:emb}
The spacetime that approximates the two-body problem can be written as a test body of reduced mass in a gravitational potential embedded in two possible background metrics:

(i) \underline{De Sitter-Schwarzschild (dSS) metric} takes a spherical-symmetric form where only $\Lambda$ drives the expansion: $ds^2 =-(1-\Phi)\,dt^{2}+\frac{dr^{2}}{1-\Phi}+r^{2}d \Omega^2\,$ with the potential $\Phi = 2 G M/(rc^2) + \Lambda r^2/6$. A test particle in the dSS metric at the low-energy limit ($\Phi \ll 1$) obeys the equation of motion (EoM):
\begin{equation}
\ddot r/r =-G M/r^3 +  \left[l/r^2\right]^2+ \Omega_\Lambda H_0^2 ,
\label{eq:kep}
\end{equation}
where $r$ is the separation, $M$ is the total mass of the binary system, $G$ is the Newtonian gravitational constant, and $c$ is the speed of light, while $l$ is the conserved angular momentum over the reduced mass. Here, $\Lambda$ is connected to the cosmological parameters via: $ \Lambda c^2 = 3\Omega_\Lambda H_0^2$, where $\Omega_\Lambda$ is the dark energy rate and $H_0$ is the Hubble parameter, which is $ 67.4\pm 0.5\,km/s/Mpc$ from the Planck collaboration~\citep{Planck:2018vyg}.

\begin{figure}[t!]
    \centering
\includegraphics[width=0.45\textwidth]{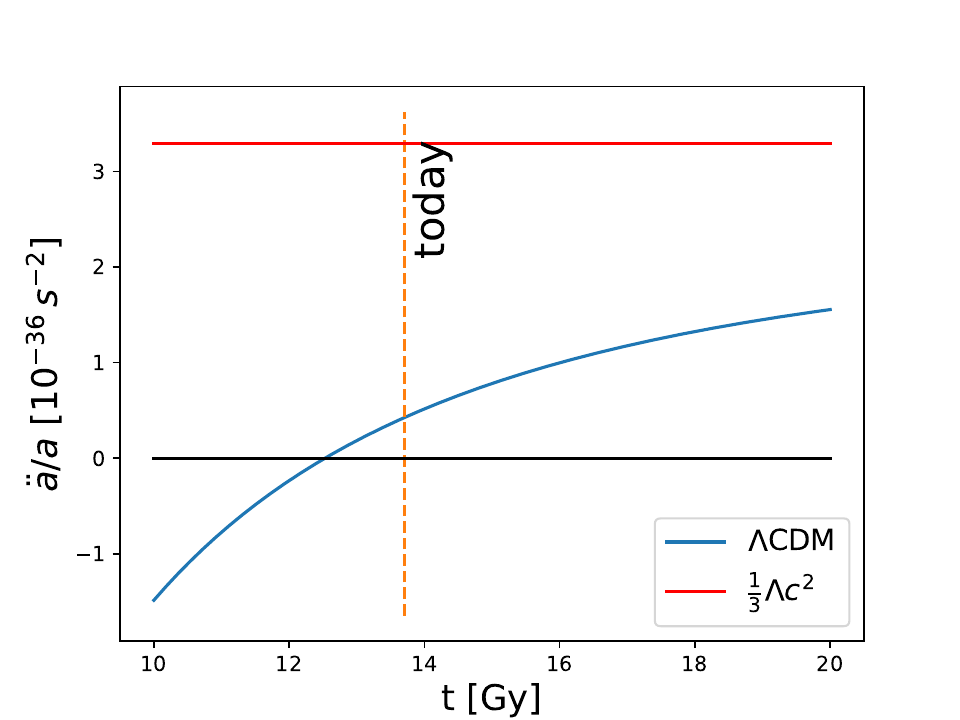}
\caption{Ratio between $\ddot{a}/a$ and $\Lambda c^2/3$ vs. time assuming the $\Lambda$CDM model. Asymptotically, both models approach the same value. For the $\ddot{a}/a$ case, the expansion rate of the Universe creates an attractive force at early times and only becomes repulsive at late times.}
\label{fig:differences}
\end{figure}

(ii) \underline{McVittie (McV) metric} describes a massive object immersed in an expanding cosmological spacetime~\citep{McVittie:1933,Kaloper:2010ec} and the full $\Lambda$ CDM model impacts the binary motion. The metric is expressed as: $ds^2 =-(1-\Phi_N)\,dt^{2}+{ a(t)^2 dr^{2}/(1-\Phi_N)}+a(t)^2 r^{2} \,d\Omega^2\,$, where $a(t)$ is the scale factor of the universe and $\Phi_N = 2 G M/(rc^2)$. In the low-energy limit ($\Phi_N \ll 1$ and $r H(t) \ll 1$), the equation of motion is expressed as~\citep{Sereno:2007tt,Faraoni:2007es,Nandra:2011ug}:
\begin{equation}
\ddot r/r =-G M/r^3 +  \left[l/r^2\right]^2 + \ddot{a}/a.
\label{eq:eomDaDt}
\end{equation}
The main difference between the two descriptions is, thus, the $\Lambda c^2/3$ vs. $\ddot{a}/a$ term in Eqs.~(\ref{eq:kep}) and (\ref{eq:eomDaDt}). Figure~(\ref{fig:differences}) displays the ratio between these two contributions over the cosmic time, assuming the $\Lambda$CDM model. While $\Lambda c^2/3$ remains constant, for $z > 0.67,$ the ratio of $\ddot{a}/a$ acts as an attractive force. This work shows that these two different embedding yield different LG masses. Thus, only based on the LG mass and dynamics, we show that  $\ddot{a}/a$ describes the LG history more consistently. The critical time for the dSS is about $35 Gys$ and for the McV is the age of the Universe~\cite{Benisty:2023clf}.

The peculiar motion of M31 is added to the Hubble expansion via the known relations: $v_{\text{LoS}} = v_{pec} \hat{r}_0 + H_0 r_0$, where $v_{\text{LoS}}$ is the line of sight velocity, $v_{pec}$ is the projected peculiar velocity on the line of sight direction, and $r_0$ is the distance from MW to M31. This relation could be obtained by integrating Eq.~(\ref{eq:eomDaDt}) over time, which shows that $\ddot{a}/a$ is consistent from a theoretical point of view. However, to describe the complete motion of the M31, it is necessary to include the tangential velocity.

\begin{figure}[t!]
\centering
\includegraphics[width=0.4\textwidth]{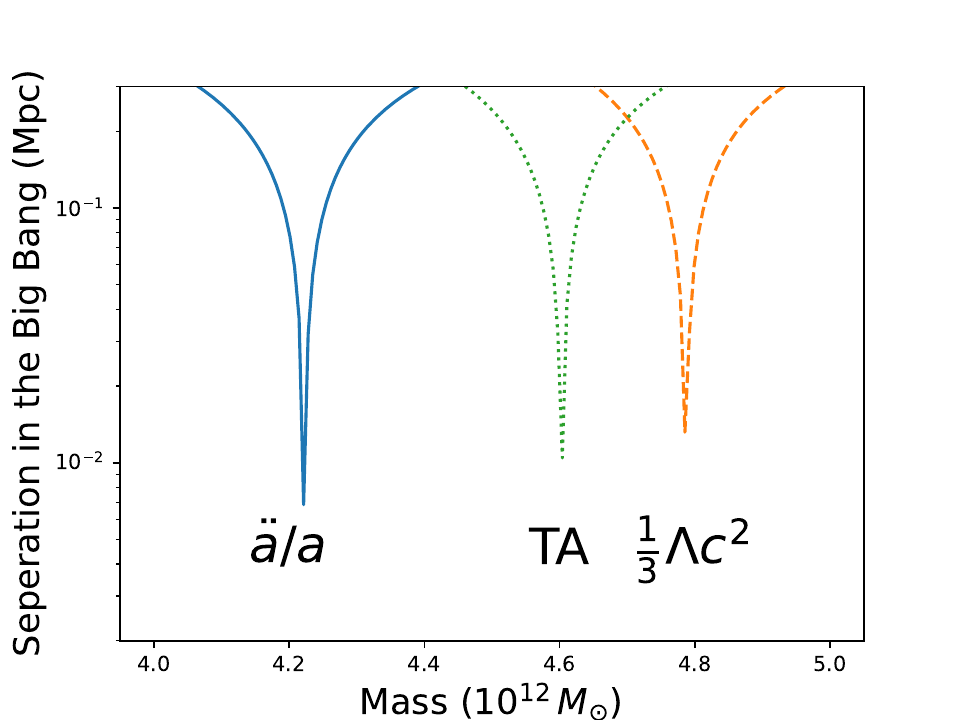}
\caption{Separation at the Big Bang between MW and M31 for different forces. The minima identify the TA masses, since they satisfy the condition of $r\left[t = 0\right] = 0$. The TA mass for the $\Lambda$ embedding is higher mass then the Newtonian TA, and the $\ddot{a}/a$ embedded mass is the lowest one. } 
\label{fig:differencesMasses} 
\includegraphics[width=0.48\textwidth]{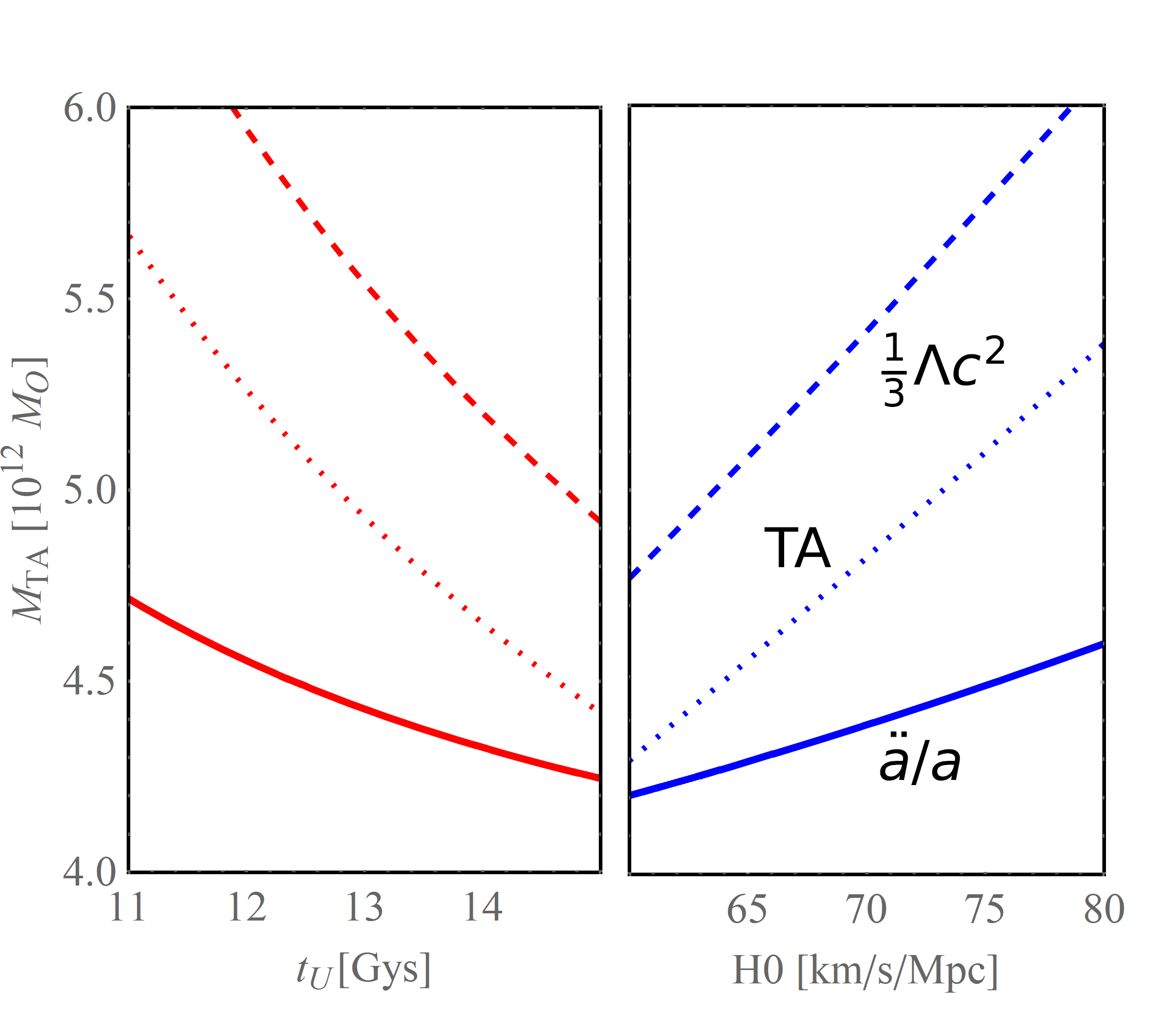}
\caption{Dependence between the predicted TA mass vs the age of the Universe (left) and vs the value of the Hubble parameter $H_0$ (right), with the mean values of the separation and the radial velocity. For larger Universe ages, the predicted TA mass is lower and for larger $H_0$ the TA mass is higher. The TA +$\Lambda$ mass is always the highest and the TA + $ \ddot{a}/a$ mass is the lowest one for any $H_0$. }
\label{fig:ageH0} 
\end{figure}

\section{Timing argument (TA) and other effects}
\label{sec:effects}
\subsection{Timing argument mass}
The TA mass is numerically calculated for the separation and the chosen velocities. We can numerically solve Eq.~(\ref{eq:kep}) or Eq.~(\ref{eq:eomDaDt}) by integrating backwards in time to the Big Bang, where the age of the Universe for $\Lambda$CDM is:
\begin{equation}
\begin{split}
t_U = \frac{2}{3H_0 \sqrt{\Omega_\Lambda}}\sinh^{-1}\sqrt{\frac{\Omega_\Lambda}{1-\Omega_\Lambda}}\;.
\label{eq:age}
\end{split}
\end{equation}
The separation and the radial velocity of M31 from the MW are known from~\citep{vanderMarel:2012xp}: $r_0 = 0.77 \pm 0.04\, Mpc$ and $v_{rad} =  109.3 \pm 4.4\, km/sec$. However, there are different inferred tangential velocities based on the galaxies proper motion (PM): Using only HST PM, \cite{vanderMarel:2012xp} reported $v_{tan} = 17 \pm 17 km/s$, which implies the classical infall model. Later, combining HST and Gaia DR2 PM, \cite{derMarel:2019} found a somewhat higher value and measure $v_{tan} = 57^{+35}_{-31} km/s$. This measurement includes a weighted average of an HST-only value of $v_{tan} = 36^{+39}_{-26} km/s$.~\cite{Salomon:10.1093} measure d$v_{tan} = 82.4 \pm 31.2 km/s$ using Gaia-DR3 PM of blue main-sequence stars in M31. We chose the value of $v_{tan} = 57 \pm 35 km/s$ because it is similar to the combined result of HST and Gaia, as in~\cite{derMarel:2019}.

\begin{figure}
\includegraphics[width=0.45\textwidth]{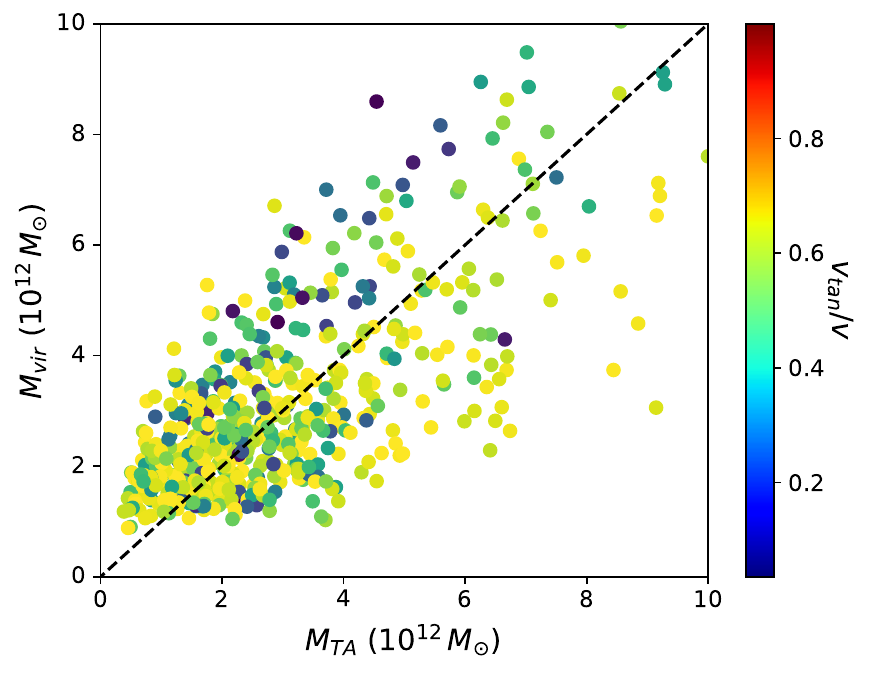}
\caption{Correlation between the TA mass and the simulation based mass. These $\sim 600$ pairs are selected from the IllustrisTNG cosmological simulation, in which the mass correction factor is computed as the ratio of the virilized halo mass of both galaxies to the mass obtained by the Newtonian TA. The mass ratio is expressed as: $M_{\text{vir}}/M_{\text{TA}} = 0.82 \pm 0.20$.}
\label{fig:CBSim}
\end{figure}

Figure~(\ref{fig:differencesMasses}) shows the separation of LG-like systems for different background models: a purely Newtonian model, adding the cosmological constant correction $\Lambda$ (dSS) and the embedding in an expanding $\Lambda$CDM Universe $\ddot{a}/a$ (McV). The best solution for minimal separation indicates the TA mass $M_{TA}$. The predicted TA mass from $\Lambda$ ($\sim 5 \cdot 10^{12} \, M_{\odot}$) is higher than the one coming from the purely Newtonian model ($\sim 4.6 \cdot 10^{12} \, M_{\odot}$), since it's necessary to overcome the repulsion force of $\Lambda$ via stronger Newtonian attraction. The $\ddot{a}/a$ model yields the lowest mass ($\sim 4 \cdot 10^{12} \, M_{\odot}$), since the $\ddot{a}/a$ acts as an attractive force at early times, thus less gravitational attraction (via mass) is needed.

The Hubble parameter enters both in the propagation of the separation at the Big Bang (the age of the Universe from Eq.~(\ref{eq:age})) and also for the force coming from the embedding. Figure~(\ref{fig:ageH0}) shows the dependence between the predicted mass based on the TA versus the age of the Universe (left) and versus the value of the Hubble parameter $H_0$ (right). For a higher age of the  Universe, the predicted TA mass is lower, and for larger $H_0$ the TA mass is higher. For all of these cases, the TA+$\Lambda$ mass is the highest and the TA$+\ddot{a}/a$ mass is the lowest one. For the same mass, the age of the Universe predicted by McV would be the higher one, since, systematically, the TA mass from the McV spacetime is the lowest one.

\begin{figure}
\includegraphics[width=0.45\textwidth]{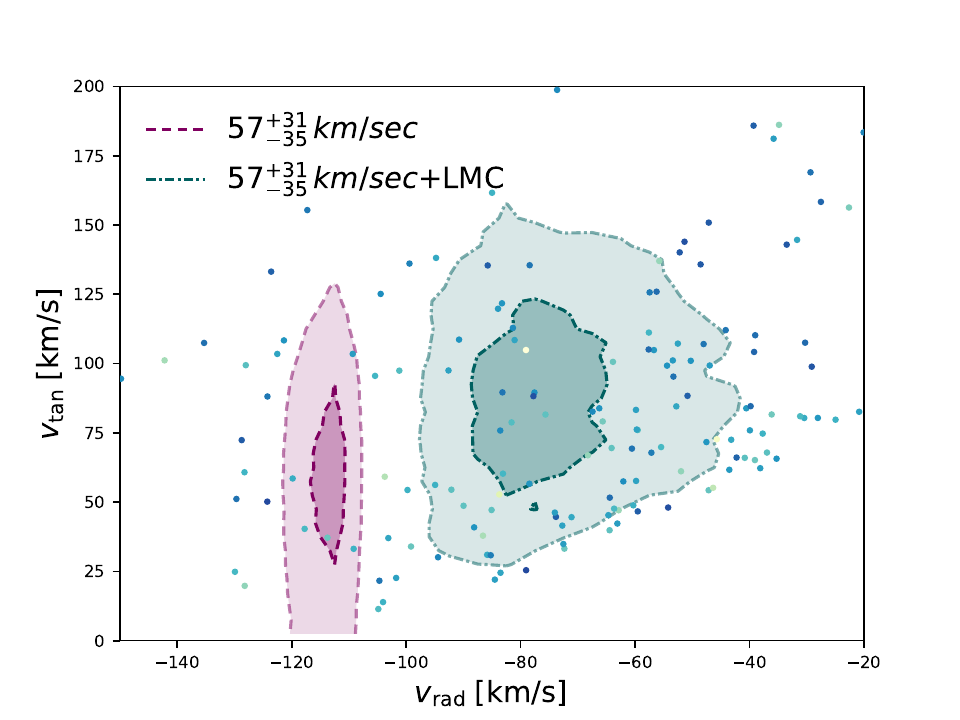}
\caption{Posterior distribution of values of M31's radial and tangential velocity for the Monte Carlo samples generated from the prior on the magnitude of $v_{tan}$ from~\citep{derMarel:2019} (purple). After compensating for the LMC perturbation, \citep{Benisty:2022ive} generates the corresponding $v_{rad}$ and $v_{tan}$ that used here as the modified effect of the LMC (green). The $v_{rad}$ is increased by $[25,30\,km/s]$ and $v_{rad}$ is shifted to $-75 \pm 15 km/s$.}
\label{fig:LMCImpact}
\end{figure}

\subsection{Simulation calibration} 
Owing to simplifications in the TA, the mass estimate may suffer from systematic bias and scatter, for instance: (i) assuming an isolated system with constant total mass, which need not be the case in reality, and (ii) assuming the two galaxies as two point particles, when they are extended halo objects with internal structure dynamics. \cite{Li:2007eg} found that the TA mass is unbiased, although it does contain some scatter, by using analogues of the LG pair from the Millenium simulation. However, \cite{Gonzalez:2013pqa} noted that the TA mass is only unbiased on average and may be an overestimation if the pairs are restricted to have similar radial and tangential velocities as the virialized MW and M31 masses.

Therefore, we used the IllustrisTNG N-body and hydrodynamical simulation calibrated by~\cite{Hartl:2021aio} to address these additional effects, because \cite{Hartl:2021aio} found a tendency for the TA mass to be overestimated. We use their sample of $\sim 600$ pairs and by imposing the cuts: (i) $r \in [650,950] \,$kpc; (ii) a mass ratio within a factor of 4; (iii) $v_{r} \in [-200,-50] \,km/s$; and (iv) $v_{\text{tan}} \in [50,200] \,km/s$, so that it resembles the observed distribution of the LG. Figure~\ref{fig:CBSim} displays the correlation between the TA mass and the simulation-based mass, with a ratio of $\text{P}(M_{vir}/M_{TA}) = 0.82 \pm 0.20$. 
For any $r,v_{rad},v_{tan}$, we identified the closest matches from the simulated pairs and identified their correction factor, $M_{vir}/M_{TA}$, to probe the total virialized mass of the galaxies, $M_{vir} = M_{MW} + M_{31}$. Including the calibration of the selected simulated pairs yields a lower mass, namely: $\left(3.89 \pm 0.62\right) \cdot 10^{12} M_{\odot}$.

\begin{figure*}[t!]
\centering
\includegraphics[width=0.9\textwidth]{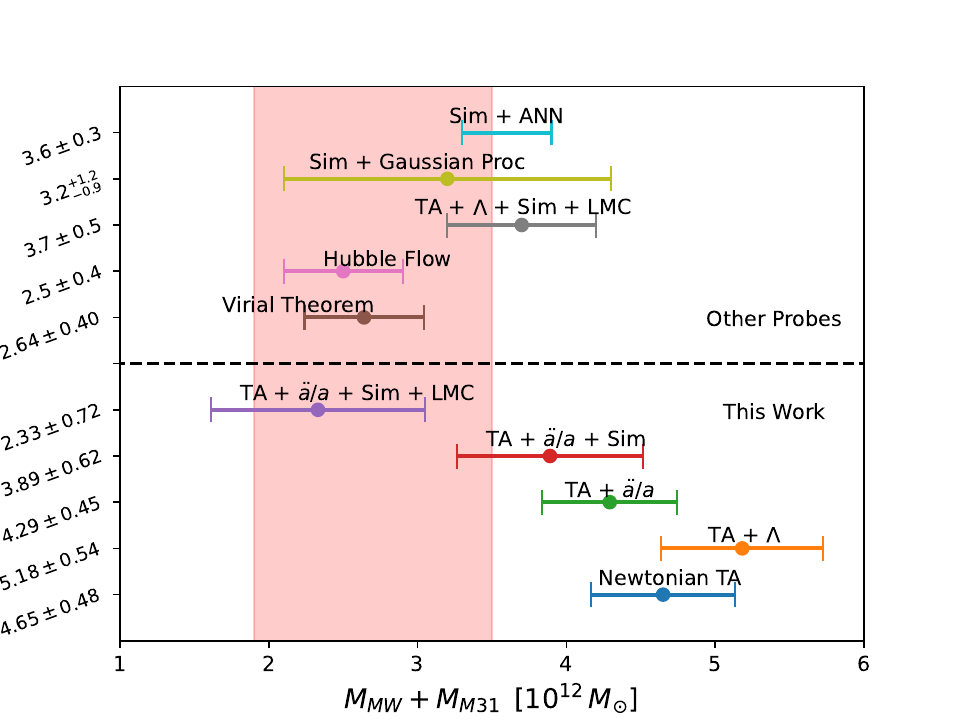}\caption{Predicted total mass of the LG of galaxies based on the proper motion of M31 towards us (TA) with the simulation calibrations. The figure compares masses inferred from different embedding with $\ddot{a}/ a$ vs. pure $\Lambda$, and from different methods in the literature: the virial theorem~\citep{Diaz:2014kqa},  Hubble flow ~\citep{Penarrubia:2014oda},  TA + $\Lambda$ + LMC~\citep{Benisty:2022ive}, artificial neural networks on simulated pairs~\citep{2017JCAP...12..034M}, and Gaussian processes on simulated pairs~\citep{Sawala:2022ayk}. The red area shows the uniform range $M_{MW} + M_{31} \in [1.9,3.3] \cdot 10^{12}\, M_{\odot}$ which is based on complementary total mass estimate of the MW and M31 galaxies~\citep{Wang:2019ubx,Sawala:2022ayk}. The total LG virialized mass should be higher then the total halo masses. }
\label{fig:mass}  
\end{figure*}

\subsection{Impact of the Large Magellanic Cloud (LMC)}
The LMC attracts the center of the MW and perturbs the LG motion. The central region of the MW, including the solar neighborhood, is drawn downward toward the LMC during its pericenter passage, while the outer, slower moving parts remain stationary. Consequently, the observed heliocentric line-of-sight velocity and PM of M31 are influenced by this downward pull. We aim to correct for this effect.

\citet{Penarrubia:2015hqa} addressed the LMC's influence by modeling it as a two-point-mass system with the MW assuming that M31 orbits the barycenter of this combined system. They determined the displacement and velocity shift of the MW relative to the barycenter by multiplying the relative position and velocity of the LMC in the Milky Way-centered frame by the mass ratio of between the LMC mass to the total mass. These corrections need to be subtracted from M31's current position and velocity in the Milky Way-centered frame before calculating its trajectory in the barycentric system.

Based on the method that was proposed in~\cite{2022MNRAS.511.2610C}, \cite{Benisty:2022ive} reevaluated the TA+Sim calibration for the LG mass. This approach first determines the past trajectories of the MW and the LMC under their mutual gravitational influence, using the distance-dependent forces from each galaxy. Then, the orbit of M31 is integrated backward in time within this time-dependent potential until the LMC perturbation is negligible. From this point, the motion of the binary system only containing MW and M31 without the LMC is propagated forward to the present time. The resulting separation and velocities of this pure binary system then yield another mass estimate via the TA and the simulation calibration. 

Figure~\ref{fig:LMCImpact} shows the posterior distribution of M31's Galactocentric radial and the velocity components $v_{rad},v_{tan}$, with and without the LMC correction generated in~\citep{Benisty:2022ive}. Using a prior from \citep{derMarel:2019} results in $v_{tan} = 59 \pm 34\,km/s$. Correcting for the LMC perturbation increases $v_{tan}$ by $[25-30]~km/s$, changes $v_{rad}$ from $-114\pm 1\,km/s$ to $-75\pm 15 \, km/s$; it also increases the distance by approximately 40 kpc. Using the modified (r,$v_{rad}$,$v_{tan}$) posterior to the LMC change, with the simulation calibration, yields a systematically lower mass $\left(2.33 \pm 0.72\right) \cdot 10^{12} M_{\odot}$. For a comparison, the combination for the TA mass with $\Lambda$ with the LMC and the simulation calibration yields higher predicted mass $3.1^{+1.3}_{-1.0} \cdot 10^{12}\,M_{\odot}$~\cite{Benisty:2022ive}; however, when including the $\ddot{a}/a$ embedding instead, we get a mass which is around $\sim 2 \cdot 10^{12} \, M_{\odot}$.

\section{Discussion:  LG Mass}
\label{sec:lgmass}
Different equations of motion and different histories produce different LG masses. Fig.~(\ref{fig:mass}) shows the different predicted masses for the different histories. The $\ddot{a}/a$ model gives the lowest mass and embedding of $\Lambda$ requires the largest mass.

Simulations of galaxy pairs predict a lower mass since the simulations take into account the extent of the galaxy halo density instead of assuming point particles~\citep{Sawala:2022ayk}. For example, TA predicts $[4,5] \cdot 10^{12}\, M_\odot$ and machine learning approaches in simulated pairs predict $\sim 3 \cdot 10^{12} \cdot M_{\odot}$~\citep{2017JCAP...12..034M,Sawala:2022ayk}. One exceptional case is an analysis with a likelihood-free approach, which predicts a broader range for the mass: $4.6^{+2.3}_{-1.8}\cdot 10^{12}\, M_{\odot}$ with large errors~\citep{Lemos:2020vhj}. By choosing the ten closest matches from the simulated pairs to the data, the predicted mass reduces to $\sim 3.5\cdot10^{12}\,M_{\odot}$. This value is compatible with the machine learning approaches. Including the calibration of the selected simulated pairs yields a lower mass $\left(3.89 \pm 0.62\right) \cdot 10^{12} M_{\odot}$.

The TA assumes the total galaxy masses to be constant, while simulations hint at growing galaxy masses over time. There are different studies in the literature to deal with this issue: (i) The Hubble flow method infers the mass by fitting the asymptotic solution of the dwarf galaxies to the velocity distance relation~\cite{Penarrubia:2014oda}. (ii) Ref.~\cite{Sawala:2023sec} claims that the total enclosed mass between the galaxy pair is invariant over a long history. (iii) In our analysis, the simulated pairs correct for the possible time dependence of the galaxy masses.

The LMC reduces the predicted mass due to the shift of the MW to about $\sim 2\cdot 10^{12}\,M_{\odot}$, as noted in \cite{Penarrubia:2015hqa,Benisty:2022ive}. Based on the method that was proposed in~\cite{2022MNRAS.511.2610C}. \cite{Benisty:2022ive} re-evaluated the LG initial conditions with the LMC, by determining the past trajectories of the MW under the gravitational influence of the LMC, integrating M31 backwards in time within this time-dependent potential until the LMC perturbation is negligible and propagating it forward to the present time. After this double propagation, we get analogous separation and velocities that simulated the MW-M31 without the LMC. The TA for analogous separation and velocities with simulation calibration yields a lower mass $\left(2.33 \pm 0.72\right) \cdot 10^{12} M_{\odot}$. For comparison, ~\cite{Benisty:2022ive} considered the effects of $\Lambda$ + Sim + LMC and obtained $3.1^{+1.3}_{-1.0} \cdot 10^{12}\,M_{\odot}$. In contrast, including $\ddot{a}/a$ reduces the mass to lower values close to $\sim 2\cdot 10^{12}\,M_{\odot}$.

This result is compatible with other probes in the literature, which are biased low: (i) the virial theorem of the MW-M31 system with its dwarf galaxies~\citep{Diaz:2014kqa,Hartl:2021aio} or (ii) fitting the motions of the dwarf galaxies to the Hubble flow ~\citep{Peirani:2005ti,Peirani:2008qs,Karachentsev:2008st,Penarrubia:2014oda,Teerikorpi:2010zz,DelPopolo:2022sev}. The result in this paper reaches a comparable mass only from the MW-M31-LMC, independently of the motions of dwarf galaxies.

The TA is biased high compared to other estimates (see Fig~\ref{fig:mass}). \cite{Sawala:2023sec} interpreted the additional mass as the dark matter in between the galaxies motivating by the N-body simulation. In our case, the combination $\ddot{a}/a$ + Sim + LMC yields a lower mass estimate, which is compatible with ~\cite{Diaz:2014kqa,Penarrubia:2014oda,Penarrubia:2015hqa}, without accounting for the possible dark matter around the halos. Hence, further analyses to constrain the (not directly detectable) dark matter are required to resolve this issue.

We compared our total mass estimates with the sum of the individual galaxy masses. There are several values for the separate masses of each galaxy in the literature: $M_{MW} \in [0.9,1.3] \cdot 10^{12}\, M_{\odot}$~\citep{Wang:2019ubx} and $M_{31} \in [1.,2.] \cdot 10^{12}\, M_{\odot}$~\citep{Sawala:2022ayk} and references therein. The red area in Fig.~(\ref{fig:mass}) shows the uniform range $M_{MW} + M_{31} \in [1.9,3.3] \cdot 10^{12}\, M_{\odot}$. The mass of the LG has to be at least in that range; higher values could arise because of the unknown dark matter distributed around the galaxies. The masses via the virial theorem and the Hubble flow (which are biased low) are within the total mass estimate. In an earlier study, Ref.~\cite{Benisty:2022ive} used the TA in a $\Lambda$-embedding with the simulation calibration and the impact of the LMC and obtained a mass, which falls into the red area, but is biased high. In this research, we changed the embedding from $\Lambda$ to the full $\ddot{a}/a$ in a $\Lambda$CDM universe and found a stronger compatibility, with these independent probes. For the first time, we see that such a low prediction for the LG mass also emerges from the TA. Further investigations to map the dark matter distribution are required to reconcile all mass estimates to a higher precision and accuracy.

\begin{acknowledgements}
DB thanks Jorge Pe\~narubbia, Jenny Wagner and Noam Libeskind for useful discussions and suggestions. DB thanks for Carl-Wilhelm Fueck Stiftung and the Margarethe und Herbert Puschmann Stiftung. DB has received partial support from the European COST action CA21136.
\end{acknowledgements}

\bibliographystyle{aa} 
\bibliography{ref.bib}

\begin{thebibliography}{38}
\expandafter\ifx\csname natexlab\endcsname\relax\def\natexlab#1{#1}\fi

\bibitem[{Aghanim {et~al.}(2020)}]{Planck:2018vyg}
Aghanim, N. {et~al.} 2020, Astron. Astrophys., 641, A6, [Erratum:
  Astron.Astrophys. 652, C4 (2021)]

\bibitem[{Benisty {et~al.}(2022)Benisty, Vasiliev, Evans, Davis, Hartl, \&
  Strigari}]{Benisty:2022ive}
Benisty, D., Vasiliev, E., Evans, N.~W., {et~al.} 2022, Astrophys. J. Lett.,
  928, L5

\bibitem[{Benisty {et~al.}(2024)Benisty, Wagner, \& Staicova}]{Benisty:2023clf}
Benisty, D., Wagner, J., \& Staicova, D. 2024, Astron. Astrophys., 683, A83

\bibitem[{Chamberlain {et~al.}(2023)Chamberlain, Price-Whelan, Besla,
  Cunningham, Garavito-Camargo, Pe\~narrubia, \&
  Petersen}]{Chamberlain:2022fqr}
Chamberlain, K., Price-Whelan, A.~M., Besla, G., {et~al.} 2023, Astrophys. J.,
  942, 18

\bibitem[{Chernin {et~al.}(2015)Chernin, Emelyanov, \&
  Karachentsev}]{Chernin:2015nga}
Chernin, A.~D., Emelyanov, N.~V., \& Karachentsev, I.~D. 2015, Mon. Not. Roy.
  Astron. Soc., 449, 2069

\bibitem[{Chernin {et~al.}(2004)Chernin, Karachentsev, Valtonen, Dolgachev,
  Domozhilova, \& Makarov}]{Chernin:2003qd}
Chernin, A.~D., Karachentsev, I.~D., Valtonen, M.~J., {et~al.} 2004, Astron.
  Astrophys., 415, 19

\bibitem[{Chernin {et~al.}(2006)Chernin, Teerikorpi, \&
  Baryshev}]{Chernin:2006dy}
Chernin, A.~D., Teerikorpi, P., \& Baryshev, Y.~V. 2006, Astron. Astrophys.,
  456, 13

\bibitem[{{Chernin} {et~al.}(2009){Chernin}, {Teerikorpi}, {Valtonen}, {Byrd},
  {Dolgachev}, \& {Domozhilova}}]{chernin:2009a}
{Chernin}, A.~D., {Teerikorpi}, P., {Valtonen}, M.~J., {et~al.} 2009, arXiv
  e-prints, arXiv:0902.3871

\bibitem[{{Correa Magnus} \& {Vasiliev}(2022)}]{2022MNRAS.511.2610C}
{Correa Magnus}, L. \& {Vasiliev}, E. 2022, \mnras, 511, 2610

\bibitem[{Del~Popolo \& Chan(2022)}]{DelPopolo:2022sev}
Del~Popolo, A. \& Chan, M.~H. 2022, Astrophys. J., 926, 156

\bibitem[{Diaz {et~al.}(2014)Diaz, Koposov, Irwin, Belokurov, \&
  Evans}]{Diaz:2014kqa}
Diaz, J.~D., Koposov, S.~E., Irwin, M., Belokurov, V., \& Evans, N.~W. 2014,
  Mon. Not. Roy. Astron. Soc., 443, 1688

\bibitem[{Faraoni \& Jacques(2007)}]{Faraoni:2007es}
Faraoni, V. \& Jacques, A. 2007, Phys. Rev. D, 76, 063510

\bibitem[{Gonzalez {et~al.}(2014)Gonzalez, Kravtsov, \&
  Gnedin}]{Gonzalez:2013pqa}
Gonzalez, R.~E., Kravtsov, A.~V., \& Gnedin, N.~Y. 2014, Astrophys. J., 793, 91

\bibitem[{Hartl \& Strigari(2022)}]{Hartl:2021aio}
Hartl, O.~V. \& Strigari, L.~E. 2022, Mon. Not. Roy. Astron. Soc., 511, 6193

\bibitem[{{Kahn} \& {Woltjer}(1959)}]{bib:Kahn}
{Kahn}, F.~D. \& {Woltjer}, L. 1959, \apj, 130, 705

\bibitem[{Kaloper {et~al.}(2010)Kaloper, Kleban, \& Martin}]{Kaloper:2010ec}
Kaloper, N., Kleban, M., \& Martin, D. 2010, Phys. Rev. D, 81, 104044

\bibitem[{Karachentsev {et~al.}(2009)Karachentsev, Kashibadze, Makarov, \&
  Tully}]{Karachentsev:2008st}
Karachentsev, I.~D., Kashibadze, O.~G., Makarov, D.~I., \& Tully, R.~B. 2009,
  Mon. Not. Roy. Astron. Soc., 393, 1265

\bibitem[{Lemos {et~al.}(2021)Lemos, Jeffrey, Whiteway, Lahav, Libeskind, \&
  Hoffman}]{Lemos:2020vhj}
Lemos, P., Jeffrey, N., Whiteway, L., {et~al.} 2021, Phys. Rev. D, 103, 023009

\bibitem[{Li \& White(2008)}]{Li:2007eg}
Li, Y.-S. \& White, S. D.~M. 2008, Mon. Not. Roy. Astron. Soc., 384, 1459

\bibitem[{{Lynden-Bell}(1981)}]{bib:Lynden}
{Lynden-Bell}, D. 1981, The Observatory, 101, 111

\bibitem[{{McLeod} {et~al.}(2017){McLeod}, {Libeskind}, {Lahav}, \&
  {Hoffman}}]{2017JCAP...12..034M}
{McLeod}, M., {Libeskind}, N., {Lahav}, O., \& {Hoffman}, Y. 2017, \jcap, 2017,
  034

\bibitem[{{McVittie}(1933)}]{McVittie:1933}
{McVittie}, G.~C. 1933, \mnras, 93, 325

\bibitem[{Nandra {et~al.}(2012)Nandra, Lasenby, \& Hobson}]{Nandra:2011ug}
Nandra, R., Lasenby, A.~N., \& Hobson, M.~P. 2012, Mon. Not. Roy. Astron. Soc.,
  422, 2931

\bibitem[{Partridge {et~al.}(2013)Partridge, Lahav, \&
  Hoffman}]{Partridge:2013dsa}
Partridge, C., Lahav, O., \& Hoffman, Y. 2013, Mon. Not. Roy. Astron. Soc.,
  436, 45

\bibitem[{Pe\~narrubia {et~al.}(2016)Pe\~narrubia, G\'omez, Besla, Erkal, \&
  Ma}]{Penarrubia:2015hqa}
Pe\~narrubia, J., G\'omez, F.~A., Besla, G., Erkal, D., \& Ma, Y.-Z. 2016, Mon.
  Not. Roy. Astron. Soc., 456, L54

\bibitem[{Pe\~narrubia {et~al.}(2014)Pe\~narrubia, Ma, Walker, \&
  McConnachie}]{Penarrubia:2014oda}
Pe\~narrubia, J., Ma, Y.-Z., Walker, M.~G., \& McConnachie, A. 2014, Mon. Not.
  Roy. Astron. Soc., 443, 2204

\bibitem[{Peebles \& Ratra(2003)}]{Peebles:2002gy}
Peebles, P. J.~E. \& Ratra, B. 2003, Rev. Mod. Phys., 75, 559

\bibitem[{Peirani \& de~Freitas~Pacheco(2006)}]{Peirani:2005ti}
Peirani, S. \& de~Freitas~Pacheco, J.~A. 2006, New Astron., 11, 325

\bibitem[{Peirani \& Pacheco(2008)}]{Peirani:2008qs}
Peirani, S. \& Pacheco, J. A. D.~F. 2008, Astron. Astrophys., 488, 845

\bibitem[{{Raychaudhury} \& {Lynden-Bell}(1989)}]{1989MNRAS.240..195R}
{Raychaudhury}, S. \& {Lynden-Bell}, D. 1989, \mnras, 240, 195

\bibitem[{{Salomon} {et~al.}(2021){Salomon}, {Ibata}, {Reyl{\'e}}, {Famaey},
  {Libeskind}, {McConnachie}, \& {Hoffman}}]{Salomon:10.1093}
{Salomon}, J.~B., {Ibata}, R., {Reyl{\'e}}, C., {et~al.} 2021, \mnras, 507,
  2592

\bibitem[{Sawala {et~al.}(2023{\natexlab{a}})Sawala, Pe\~narrubia, Liao, \&
  Johansson}]{Sawala:2023sec}
Sawala, T., Pe\~narrubia, J., Liao, S., \& Johansson, P.~H. 2023{\natexlab{a}},
  Mon. Not. Roy. Astron. Soc., 526, L77

\bibitem[{Sawala {et~al.}(2023{\natexlab{b}})Sawala, Teeriaho, \&
  Johansson}]{Sawala:2022ayk}
Sawala, T., Teeriaho, M., \& Johansson, P.~H. 2023{\natexlab{b}}, Mon. Not.
  Roy. Astron. Soc., 521, 4863

\bibitem[{Sereno \& Jetzer(2007)}]{Sereno:2007tt}
Sereno, M. \& Jetzer, P. 2007, Phys. Rev. D, 75, 064031

\bibitem[{Teerikorpi \& Chernin(2010)}]{Teerikorpi:2010zz}
Teerikorpi, P. \& Chernin, A.~D. 2010, Astron. Astrophys., 516, A93

\bibitem[{van~der Marel {et~al.}(2012)van~der Marel, Fardal, Besla, Beaton,
  Sohn, Anderson, Brown, \& Guhathakurta}]{vanderMarel:2012xp}
van~der Marel, R.~P., Fardal, M., Besla, G., {et~al.} 2012, Astrophys. J., 753,
  8

\bibitem[{{van der Marel} {et~al.}(2019){van der Marel}, {Fardal}, {Sohn},
  {Patel}, {Besla}, {del Pino}, {Sahlmann}, \& {Watkins}}]{derMarel:2019}
{van der Marel}, R.~P., {Fardal}, M.~A., {Sohn}, S.~T., {et~al.} 2019, \apj,
  872, 24

\bibitem[{Wang {et~al.}(2020)Wang, Han, Cautun, Li, \& Ishigaki}]{Wang:2019ubx}
Wang, W., Han, J., Cautun, M., Li, Z., \& Ishigaki, M.~N. 2020, Sci. China
  Phys. Mech. Astron., 63, 109801

\end{thebibliography}

\end{document}